# Simulating volume-controlled invasion of a non-wetting fluid in volumetric images using basic image processing tools

Jeff T. Gostick *,[1], Jianhui Yang [2] and Edo S. Boek [3]

[1] Department of Chemical Engineering, University of Waterloo, Waterloo, ON, N2L 3G1, Canada

[2] Geoscience Research Centre, TOTAL E & P UK Limited, Westhill, Aberdeenshire, AB32 6JZ, UK

[3] Division of Chemical Engineering & Renewable Energy, School of Engineering and Materials Science, Queen Mary University of London, Bethnal Green, UK

*Corresponding author: jgostick@uwaterloo.ca

# Abstract

A new algorithm is presented for simulating volume-controlled invasion of a non-wetting phase into voxel images. This method is complementary to the traditional morphological image opening method which mimics pressure-based invasion. A key advantage of the volume-based approach is that all saturations between 0 and 1 can obtained rather than the irregularly and widely spaced saturation steps obtained by pressure-based methods. Because of the incremental increases in saturation, it becomes possible to correctly predict defending phase trapping, which is not the case when pressure-based steps are applied. The algorithm is validated against morphological image opening and obtains near perfect agreement at equal saturations as expected from theory. It is also demonstrated that a volume-controlled capillary pressure curve can be obtained that displays the characteristic jumps in capillary pressure, and moreover, the envelop of peak pressures yields the pressure-based capillary pressure obtained by morphological opening, so in fact the results of the proposed algorithm are a superset of the morphological approach. Finally, results are compared to multiphase lattice Boltzmann and qualitatively similar results were achieved in substantially less time. The lattice Boltzmann method is more flexible in terms of variable contact angle and inclusion of viscous effects, but for quasi-static volume-based injection of a non-wetting fluid, the proposed method is viable alternative.

# 1. Introduction

Volumetric images of porous materials can be obtained by various techniques including X-ray tomography (Wildenschild and Sheppard, 2013) (XCT) and focused ion-beam serial sectioning (Sabharwal et al., 2016) (FIB-SEM). It is also possible to generate images artificially (Hinebaugh et al., 2017; Mohebi et al., 2009; Mosser et al., 2017; Pant et al., 2014; Talukdar et al., 2002) for materials that cannot be easily imaged, or to study the impact of parameters that would be difficult to adjust experimentally such as pore size distributions. In any case, with such an image available it is possible to study structural properties and transport parameters of the material directly on the image (Bultreys et al., 2016; García-Salaberri et al., 2018; Islam et al., 2018; Kok et al., 2019; Starnoni et al., 2017). For instance, information about the pore size distribution can be obtained by analyzing the result of the local thickness filter or chord length distribution (Gostick et al., 2019; Petri et al., 2020; Torquato, 2005). Transport parameters such as the tortuosity factor and absolute permeability coefficient are routinely obtained by performing direct numerical simulations with the image voxels as the grid (Cooper et al., 2016; García-Salaberri et al., 2018). However, the study of multiphase flow conditions such as estimating relative permeability and diffusibility (Rosén et al., 2012; Sabharwal et al., 2018) are more challenging because they require images containing a suitable distribution of the defending and invading phases. Although it is possible to obtain X-ray tomograms on partially saturated samples (Agaesse et al., 2016; Akbarabadi et al., 2017), it is significantly more complex than on dry materials. Careful control of the applied capillary pressure is necessary, and many scans must be taken at various saturation levels to get a full picture of the evolving fluid configuration. It is therefore desirable to simulate fluid invasion into volumetric images of porous materials, thereby allowing both the simple acquisition of a single dry image, as well as providing a range of fluid distributions and saturations from which to study multiphase and capillary behavior.

There are several ways to simulate non-wetting fluid invasion into volumetric images of porous materials, each with their pros and cons. The volume of fluid method (Kohanpur et al., 2020; Raeini et al., 2012; Tranter et al., 2015) (VOF) monitors the surface between immiscible fluids, while applying a single set of momentum equations to both phases and tracking the volume fraction of both fluids in each cell of the domain. Because the fluids are immiscible, the interfaces between phases display capillary effects. The popularity of VOF is at least partly due to its availability in commercial software packages. One of the most challenging and time-consuming steps is the construction of a suitable mesh from the voxel image. Another commonly used approach is the Lattice Boltzmann method (LBM) which simulates the streaming and collision process of fictive particle groups instead of solving the Navier-Stokes equations directly. The LBM method offers the key advantage of using a uniform mesh (i.e. voxels) and can handle complex geometries without extra meshing operations. It is relatively computationally efficient and can be easily parallelized since most of the operations are local. However, LBM is relatively demanding in terms of memory requirements so cannot leverage GPUs, and limited to low Mach number incompressible flows. This is because LBM algorithms generally conserve mass and momentum but not energy (Yang et al., 2019). LBM is particularly time consuming for low velocity flows.  Phase field methods (Frank et al., 2018; Liu et al., 2019), which solve the Cahn-Hilliard equations to track diffuse interfaces, are also becoming increasingly used. This method requires solving a 4$^{th}$ order PDE, however, so is computationally intensive and complex.  Another class of invasion simulations are based on the level set method (Jettestuen et al., 2013; Prodanović and Bryant, 2006; Rezk et al., 2013), which involves the solution of a partial differential equation describing the motion of the advancing interface in geometrical terms.  This method is complex to implement, and no standard implementations are available.  Moreover, it does not easily allow for variable contact angle although this has been addressed recently (Jettestuen et al., 2013).

The physics-based simulation approaches described above offer several advantages such as adjustable wettability (i.e. contact angle or surface tensions), and incorporation of viscous effects in the invading and defending phase, which become important at higher injection rates (Lenormand et al., 1988).  However, these advantages come at a high computational cost. For instance, conducting a two-phase LBM simulation for a capillary number $Ca = 10^{-4}$ on a $500^3$ image on a desktop workstation typically requires CPU time on the order of days. Moreover, a two-phase LBM simulation at lower capillary numbers may suffer from spurious currents and therefore require smaller timesteps for a reasonable numerical stability, increasing computation time even further.  Because capillary interfaces tend to spherical shapes to minimize their surface energy, it is possible to approximate fluid configurations as an assemblage of spheres. This can be easily accomplished using image processing, which is significantly faster than the physics-based simulations mentioned above (Kemgue et al., 2019). Hazlett (Hazlett, 1995), followed by Hilpert and Miller (Hilpert and Miller, 2001) developed this approach using image morphology (Soille, 2004). In this approach, binary opening (erosion followed by dilation) of the pore space with a spherical structuring element of radius $R$ returns a Boolean mask indicating all regions of the pore space where a sphere of radius $R$ can fit. By trimming regions of the mask not connected to a specified inlet, this process mimics fluid invasion at a constant applied capillary pressure where $P_c = 2\sigma/R$. The process can be repeated at decreasing values of $R$ to obtain fluid configurations as a function of applied pressure, mimicking a drainage experiment. The main limitations of MIO approaches are: (a) The solid-fluid wettability (i.e. contact angle) cannot be varied since spherical structuring elements are used, which imply perfectly wetting/non-wetting phases (although this can be relaxed slightly by letting the structuring element extend into the solid phase (Schulz et al., 2015)); (b) throat openings with high aspect ratio are only invaded when the smallest dimension is breached since the invasion is based on *inscribed* spheres thus tending to overestimate the invasion pressure; and (c) since increased invasion saturations are obtained by increasing the pressure, large swaths of the pore space are filled in a single step due to shielding and network accessibility effects. It is not uncommon for the saturation to jump as much as 50% in a single step (Mohammadmoradi and Kantzas, 2016;

Norouzi Apourvari and Arns, 2016; Sabharwal et al., 2018). This last issue leads to two additional problems, which the present work addresses. Firstly, the invading fluid configurations are not fine-grained enough to obtain accurate multiphase transport correlations as functions of saturation. Secondly, trapping of wetting phase cannot be correctly assessed since trapping may occur at some saturation between two steps, but this cannot be computed until the following configuration has been determined and therefore some pore space will be incorrectly marked as invaded.

This work presents a new algorithm based on simple image analysis tools to leverage their computational efficiency while mimicking a *volume*-controlled invasion rather than pressure (or size) based. In other words, the proposed algorithm simulates an invasion percolation process (Wilkinson and Willemsen, 1983) while traditional MIO simulates ordinary percolation. Because the invading fluid is added in minimal volume increments, it is possible to obtain fluid configurations corresponding to any saturation between 0 and 1 (within the limits of voxel resolution). Additionally, it is also shown that capillary pressure curves can be obtained that include the jumps in capillary pressure that are characteristic of volume-controlled porosimetry (Knackstedt et al., 1998). Lastly, it is possible to identify trapped defending phase with ease as a post-processing step, and it is shown that traditional MIO underestimates trapping significantly. The algorithm presented herein is validated against the traditional MIO approach (Hazlett, 1995; Hilpert and Miller, 2001) to confirm that it produces equivalent fluid configurations at equivalent saturations. Results are also compared with 2-phase lattice Boltzmann simulations and demonstrate good qualitative agreement. All the code developed herein is available in the open-source python package PoreSpy (Gostick et al., 2019) that is freely available on Github (https://github.com/PMEAL/porespy) and is trivial to install, assuming the Scipy (Virtanen et al., 2020) stack is present, including Scikit-Image (van der Walt et al., 2014).

## 2. Implementation

### 2.1. The Image-Based Invasion Percolation Algorithm

The algorithm is referred to as image-based invasion percolation (IBIP) since it injects non-wetting phase into the domain in small, incremental steps following capillary entry conditions at each step, analogous to the classic invasion percolation algorithm performed on pore network (Wilkinson and Willemsen, 1983). The workflow is illustrated by a flow diagram in Figure 1. The first step is to obtain the distance transform (*dt*) of the pore space image (*image*) (Figure 2(b)). The locations of the inlets must be specified, which is done by creating an empty Boolean image (*inlets*) of the same shape as the pore space with non-zeros indicating the inlet voxels (Figure 2 (c)). The inlets can be on the edge of the image, which is the usual case, but could also be in the center of the image for simulating evaporation for instance.

The iterative or incremental portion of the algorithm, denoted by the dashed box in Figure 1, proceeds by dilating the *inlets* image with a spherical structuring element of radius $R = 1$ (equivalent to a cross) to find candidate voxels neighboring the inlet sites (denoted as purple in Figure 2 (e). Next the voxels of *dt* underlying the *inlets* voxels (original plus dilation) are examined (Figure 2 (f)) and the value and location of the maxima(s) are noted (Figure 2 (g)). These maxima points indicate the radii, $R$, and center point(s), $C$, of the largest sphere(s) that can be drawn in the image that are also connected to the inlets ($C =$ $[i, j]$ in 2D or $C = [i, j, k]$ in 3D). Spheres are then inserted into the image at these points to denote the invasion of non-wetting fluid[1]. This can be accomplished by dilating the voxels identified as insertion points in the previous step ($C$) with a structuring element of radius $R$ ((Figure 1(h)). It is possible that more than one location was identified in the previous step ( $C = \big[[i_1, j_1, k_1], [i_2, j_2, k_2], \ldots, [i_n, j_n, k_n]\big]$ ), so spheres would be inserted at all of them, meaning that injection occurs simultaneously from all identified locations. Inserting spheres into the image should be done in a certain way to aid the postprocessing discussed in Section 2.2. Firstly, the voxels where the spheres are inserted are given the current step number. Secondly, values should only be written into locations that have not already been invaded to avoid overwriting results from previous steps. This is illustrated in Figure 1(l) where the sphere from the second step is distinguishable from the first step by different grayscale values. With such an approach it becomes trivial to obtain any intermediate invasion configuration from a single image by finding all voxels labelled $\leq N$, where $N$ is an arbitrary invasion step. Converting the invasion step numbers to saturation values can be done as a postprocessing step as discussed in Section 2.2.1. Before iterating the process again, the identified insertion points, $C$, are added to the *inlets* image ((Figure 1(i)), and the same locations in *dt* are set to zero so they are not found in the next iteration. At this point, the process repeats by dilating the updated *inlets*

---

[1] Note that only spheres with integer radii can be inserted into a digital image so the value of $R$ must be truncated to the nearest integer

image (Figure 1(i)) and so on. The result of many iterations is shown in Figure 1(d) where the greyscale value indicates the step number at which it became occupied. A more complex example is shown in Figure 3(left), where the solid is indicated by grey, and blind or unreachable pores by black. Trapped pore space is not identified in this image, but this can be identified as a postprocessing step as discussed in Section 2.2.3. Figure 4 shows a 3D example.

## 2.2. Postprocessing the Image

### 2.2.1. Converting Invasion Steps to Saturation Values

Inserting the step number into the final image provides a single image containing the entire history of the invasion. This makes it very convenient to use a Boolean threshold to obtain the fluid configuration at a given step (i.e. $result \leq 1000$). This is illustrated in Figure 3 for steps up to 3000, in increments of 500.

Although it is straightforward to determine the saturation of a thresholded image, it is more convenient to perform the threshold based on saturation directly. It is possible to produce an image similar to Figure 3(left) with saturations instead of sequence values by iteratively thresholding the sequence values (as done in Figure 3(right)), then calculating the saturation value at each threshold and replacing the sequence values with the calculated saturation (only writing to locations that do not already have a saturation value). The resultant image, as shown in Figure 5(left), can then be thresholded as $\leq s$ to obtain the invading fluid configuration at any desired saturation, $s$.

### 2.2.2. Computing Capillary Pressure Curves

When performing the algorithm as laid out in Figure 1 it is possible to label the *result* image with the size of the sphere being inserted, rather than the step number (or optionally, both images can be produced simultaneously). The result is shown in Figure 5(right). This image is not suitable for thresholding, but it does reveal the regions of the image that are most easily invaded (by high greyscale value), so could be useful for quantitatively identifying spatial correlations in the pore structure, for instance. More interestingly, this image can be analyzed in combination with the saturation image shown in Figure 5(left) to produce the capillary pressure curve for the material, analogous to the result of morphological image opening. The saturation image is thresholded at increasing values of saturation, at each step noting the size of the spheres that were inserted at the newly invaded voxels. This size value can then be converted to a capillary pressure using the Washburn equation ($P_c = 2\sigma/r$) and stored in an array alongside its corresponding saturation value. The resulting curve is illustrated in Figure 6 as the green line. The horizontal noise is expected for a volume-controlled porosimetry experiment (Knackstedt et al., 1998; Sygouni et al., 2006). Once a fluid meniscus has breached a constriction, it expands to fill the large cavity beyond and hence the capillary pressure of the entire system drops. The process is repeated for each new constriction. Although most constrictions are larger, occasionally a constriction is reached that is smaller than any seen previously, so a new peak capillary pressure is required to breach it. The envelop of peak pressures corresponds to the steps in the pressure-based curve. The peak pressure is indicated by the red line in Figure 6, which was obtained using standard MIO which emulates a pressure-based porosimetry experiment. Note that the comparison in Figure 6 is only possible when trapping is not considered, since the pressure-based MIO technique cannot correctly predict trapping, as discussed in Section 3.1.1. A capillary pressure curve can be produced from IBIP using the same method described above but after assessing trapping as described in the next section. In addition to the limitations of image-based invasion mentioned in the introduction, the proposed algorithm has one additional limitation, since existing fluid interfaces do not contract or deflate when the capillary pressure of the invading phase decreases after an invasion event.

### 2.2.3. Identifying Trapped Wetting Phase

In many cases the wetting phase can become disconnected from the outlet. For example, in the case of air invading into a water-wet packing, the wetting films of water may evaporate, eliminating any pathway for water to escape creating disconnected or trapped water clusters. It is possible with the present method to identify trapped clusters of defending phase as a post-processing step. The image containing the invasion sequence values can be analyzed by thresholding sequence (or saturation) values in decreasing order and identifying any voxels invaded above the threshold value that are not connected to the outlet. This approach is similar to that of Masson (Masson, 2016) who applied such a reverse search on a pore network model. Figure 7(left) shows the result of a 2D image without removing the trapped clusters, while Figure 7(right) shows the invasion pattern with the trapped clusters set to uninvaded (denoted by gray). The color bars in Figure 7 indicate the sequence at which each pixel was invaded. The sequence numbers for the trapped case have been recalculated after the trapped voxels were set to uninvaded by re-ranking the sequence values to make them contiguous. Converting these sequence values to saturation can be done in the same manner described in Section 2.2.1 for the non-trapping case.

## 2.3. Speed Considerations

The IBIP algorithm consists of several steps, two of which are particularly time consuming: the dilation of the boundary image (*inlets*) to find insertion points (Figure 2 e and i) and the dilation of the insertion points to create spheres (Figure 2 h and l). Since the complete invasion process can require thousands of steps, applying morphological filters on the entire image at each step can become infeasible, especially for 3D images. The first step, dilating the *inlets* image, is fortunately done with a very small structuring element ($R \leq 3$) so is relatively fast. Because of this, the standard implementation in *scipy.ndimage* outperforms the parallelized versions, either the FFT-based version or the domain decomposition approach (both available in *PoreSpy*). The second step requires inserting spheres of arbitrarily large size at the identified insertion points. In the basic implementation described above this was done using a morphological dilation with the structuring element radius taken from the values of the distance transform at the insertion points. As the structuring element gets larger however, this process becomes increasingly time consuming. It was found that the following "brute-force" approach was actually fastest, which avoids processing the entire image: obtaining the $[i, j, k]$ coordinates of the insertion points then explicitly inserting spheres directly into the image at each point. Finding insertion point coordinates becomes the most time-consuming step (i.e. *numpy.where*), but it is substantially faster than applying a filter to the entire image. It was explored whether increasing the amount of dilation of the *inlets* on each step could help by identifying more voxels on each step, but this hurt the accuracy without giving a useful speed-up so was abandoned.

## 2.4. Validation

### 2.4.1. Morphological Image Opening

Hazlett (Hazlett, 1995) and Hilpert and Miller(Hilpert and Miller, 2001) proposed a means of simulating pressure-based drainage using image-processing techniques based on image morphology (Soille, 2004). Image opening is a morphological operation which finds regions of the image foreground (void space) where a sphere of a given radius can fit. Image opening is accomplished by first eroding the image by the given structuring element, then dilating the surviving voxels by the same element. This can be performed with successively smaller structuring elements to simulate pressure-based invasion of a non-wetting fluid by trimming all voxels that are disconnected from the inlet after the erosion step. This ensures that the invading clusters are always connected to the inlets. Note that the trimming must actually occur after the erosion but before the dilation otherwise bulging menisci can be connected though uninvaded throat constrictions.

Mohammadmoradi and Kantzas (Mohammadmoradi and Kantzas, 2016) refer to this as an immature pathway. MIO simulates a mercury intrusion porosimetry experiment by assuming that pore sizes can be related to capillary pressures using some geometric relationship such as the Washburn equation ($P_c = -2\sigma \cos(\theta)/R$). This process was utilized in the present work to obtain invading fluid configurations for comparison to the proposed invasion percolation approach.

### 2.4.2. Multiphase Lattice Boltzmann

A color-gradient based multi-relaxation time lattice-Boltzmann method was used to simulate the non-wetting phase invasion percolation process (drainage) (Ahrenholz et al., 2008; Yang and Boek, 2013). This method has a better numerical stability than the Shan-Chen and Free Energy models and has been widely used in two-phase flow studies. The density and viscosity of both wetting and non-wetting phase are set to be equal (1 and 0.1 in lattice units [l.u.] respectively) for optimal numerical stability. A surface tension of 0.05 [l.u.] and a pressure gradient of $5 \times^{-5}$ [l.u.] is used to achieve a capillary number between $1 \times 10^{-4}$ and $1 \times 10^{-5}$. A lower capillary number is possible but will make the simulation more time consuming. A buffer region was added to the inlet edge of the domain to allow the invading phase to stay connected to a reservoir of the same pressure. A similar buffer region is added to the outlet edge of the domain to allow both phases to pass freely to the outlet. A fixed pressure boundary condition is applied at the inlet and outlet. A recoloring operation is applied to keep the fluid entering the left buffer region as non-wetting phase.

# 3. Results and Discussion

In this section the image-based IP method is first compared to morphological opening in terms of accuracy of predicted fluid configurations, then to multiphase LBM for accuracy and computational cost.

## 3.1. Validation and Comparison

### 3.1.1. Comparison to MIO

Pressure-controlled injection is a subset of rate-controlled injection, in that the points that constitute the envelop of peak pressures observed during rate-controlled injection (IBIP) yield the data points for a pressure-controlled injection (MIO). As such, the fluid configurations at equal saturations should match exactly, and this was found to be the case here. Figure 8(left) shows a comparison of the two algorithms obtained by thresholding the MIO result at each value of saturation that it achieved, then thresholding the IBIP result at the same saturation value and comparing differences across all saturations. The yellow pixels in Figure 8(left) indicate locations that were invaded by MIO before IBIP, purple represents the reverse, and green represents pixels that were invaded simultaneously in both images. To quantify these differences properly, Figure 8(right) shows a plot of the absolute number of pixels that are in disagreement between the two images (both purple and yellow) as a function of saturation. The maximum deviation is only 24 pixels in a $500^2$ image (250,000 pixels). It is also noteworthy that the error does not accumulate between steps. These minor differences are not due to any fundamental problems with either algorithm, but to the floating-point accuracy with which the saturation was computed between images. This effect is illustrated in Figure 9(right) where the saturation values were rounded to 4 decimal places before being compared, and the absolute differences are noticeably less. Rounding even further, however, increases error as the saturation values begin to lose fidelity.

Because the MIO approach fills large fractions of the void space at each applied pressure step, it is incapable of accurately capturing trapping. The formation of smaller trapped regions that may have occurred within each step are therefore missed. This problem is illustrated in Figure 9. The left panel shows the fluid configuration predicted by IBIP with the trapped pore space indicated by grey. The center image is the result of MIO. These results are directly compared in the right panel, where all visible yellow pixels are those that were incorrectly invaded by MIO when they should have been trapped. The ability to capture each individual pore-scale trapping event is one of the more powerful features of the IBIP algorithm.

### 3.1.2. Comparison with LBM

Figure 10 shows a side-by-side visual comparison of the two-phase LBM simulation with the proposed IBIP algorithm. The qualitative agreement in both space and time sequence are generally good, but there are clearly some differences. These can be attributed to viscous effects in the LBM simulations, which conducted at a capillary number of $Ca =$ $10^{-4}$. Performing LBM simulations at lower values of $Ca$ were problematic, since a 10x reduction in $Ca$ requires ~10x more time to compute. This underscores the value of the IBIP algorithm, which essentially corresponds to a $Ca \to 0$. Such quasi-static simulations are impractically time-consuming using LBM. The LBM simulations also employed periodic boundary conditions, but this only resulted in a small amount of invasion, visible at the top and bottom of the final image. Overall, the agreement is reasonable, indicating that the IBIP algorithm predicts the correct invasion patterns, with differences attributable to the non-zero flow rate required by the LBM approach.

# 4. Conclusions

An image-based invasion percolation (IBIP) algorithm was presented that can efficiently predict non-wetting phase invasion patterns representing volume-controlled injection. The IBIP algorithm is based entirely on basic image processing tools, and complements the well-established MIO approach, which simulates pressure-based invasion. In fact, IBIP is a superset of MIO since it was shown that the envelop of peak pressures obtained by IBIP yield the MIO results. The main benefit of the IBIP algorithm is that fluid configurations are obtained for all non-wetting phase saturations, rather than just at discrete applied pressures as in MIO, allowing the study of the evolving fluid configuration in detail.

The correspondence between the two algorithms was established by comparing fluid invasion patterns as a function of saturation, and the results were in near perfect agreement. At worst only a few dozen pixels were erroneously invaded (in an image with 250,000 pixels), but this was actually due to floating-point error in the saturation values rather than any fundamental difference in the results. Further validation of the IBIP algorithm was obtained by comparison with 2-phase lattice Boltzmann simulations. The qualitative agreement was acceptable, although a quantitative pixel-by-pixel comparison was not performed. These differences are due to the fact that image-based approach is completely capillary dominated without any viscous effects, while the LBM simulations in this study were carried out at a finite capillary number.

The result of the IBIP algorithm can be converted into a volume-controlled capillary pressure curves, complete with the noisy pressure spikes that are the hallmark of these experiments (Knackstedt et al., 1998). The envelope of the pressure spikes was shown to follow the pressure-based capillary pressure curve produced by MIO as expected (Shikhov and Arns, 2015). It is therefore possible that capillary pressure curves predicted by IBIP could be used to study the dynamics of capillary displacement, using signal processing techniques for instance (Sygouni et al., 2006). It was also shown that trapping of the defending phase can be easily applied using a post-processing step. MIO cannot truly predict trapping since large swaths of the pore space are filled in a single step, and therefore many discrete pore trapping events are not observed. IBIP by contrast, fills the image using a volume-based approach, so the growth of individual menisci is captured, and therefore trapping can be accurately assessed. Comparing the fluid configurations between MIO and IBIP shows that MIO significantly underpredicted the amount of trapped defending phase. The ability to predict trapping as well as produce images at any value of saturation makes IBIP highly appealing for studying multiphase flow properties such as relative permeability and diffusibility.

# 6. Figures and Tables

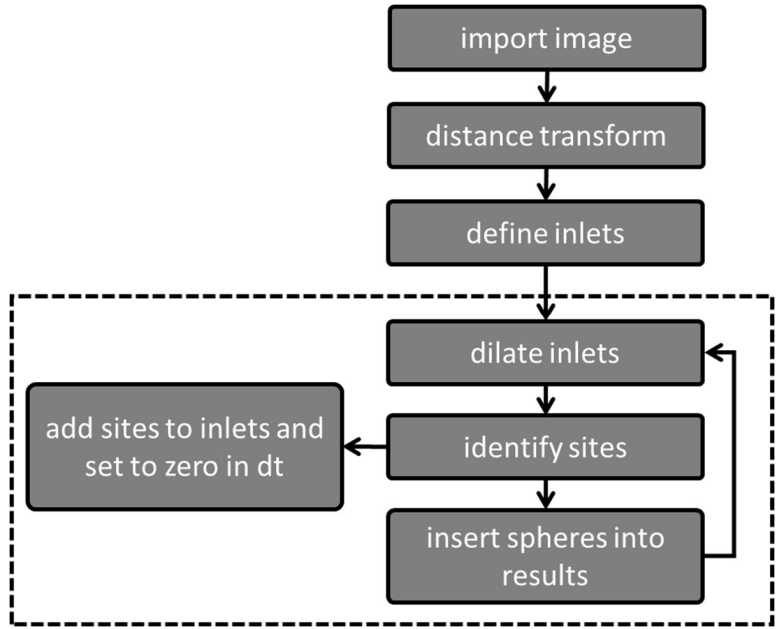

**Figure 1: Flow diagram illustrating the sequence of image processing steps**

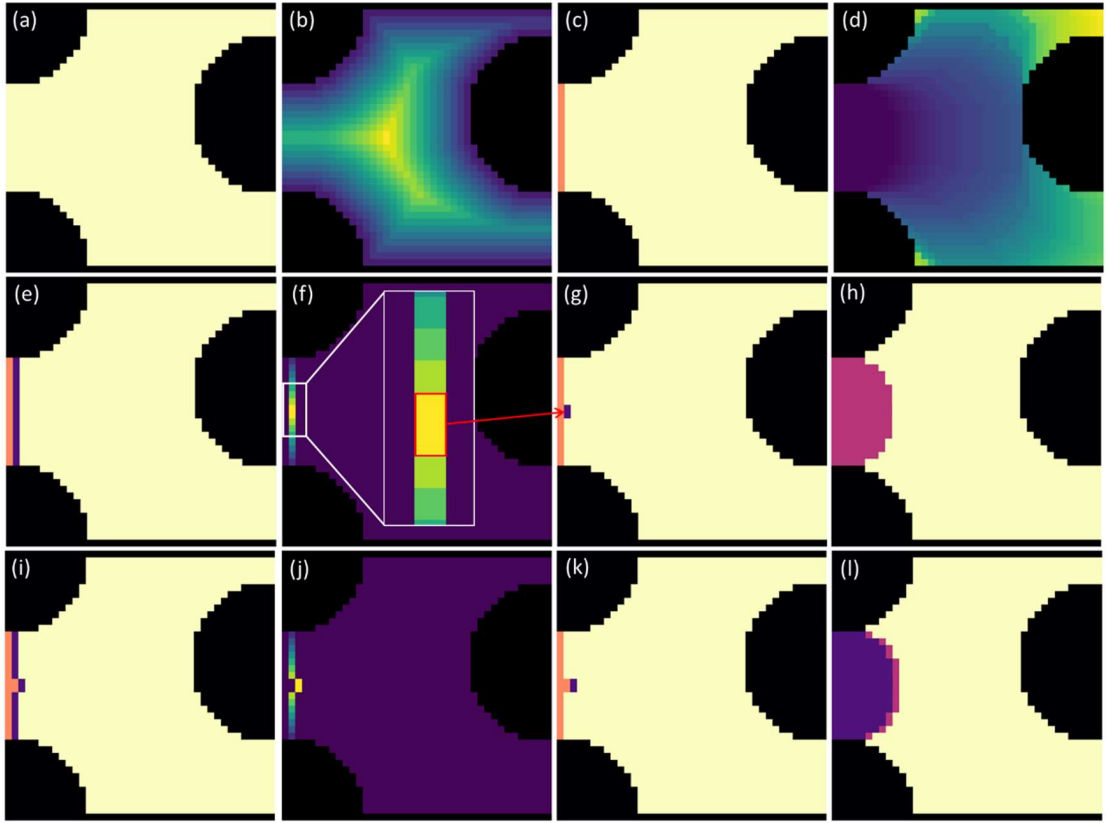

Figure 2: 2D illustration of steps involved in the invasion sequence. (a) Shows the original pore space, with yellow indicating void and black indicating solid. (b) shows the Euclidean distance transform of the void space, which is repeatedly used in subsequent steps. (c) shows the boundary image, with pink indicating the inlet voxels for the invasion. (d) shows the final result of the invasion simulation with the greyscale value indicating the step at which each pixel was invaded. (e to h) show the basic operations for a single invasion step, starting with dilation of the border (e), followed by inspection of the Euclidean distance transform underlying the dilated border (f) identification of the pixel(s) containing the maximum value (g), and lastly dilation of the identified pixel(s) by a structuring element with a radius equal to the maximum value of the masked distance transform (h). (i to l) show the process repeated, with the points identified in (g) now part of the boundary image.

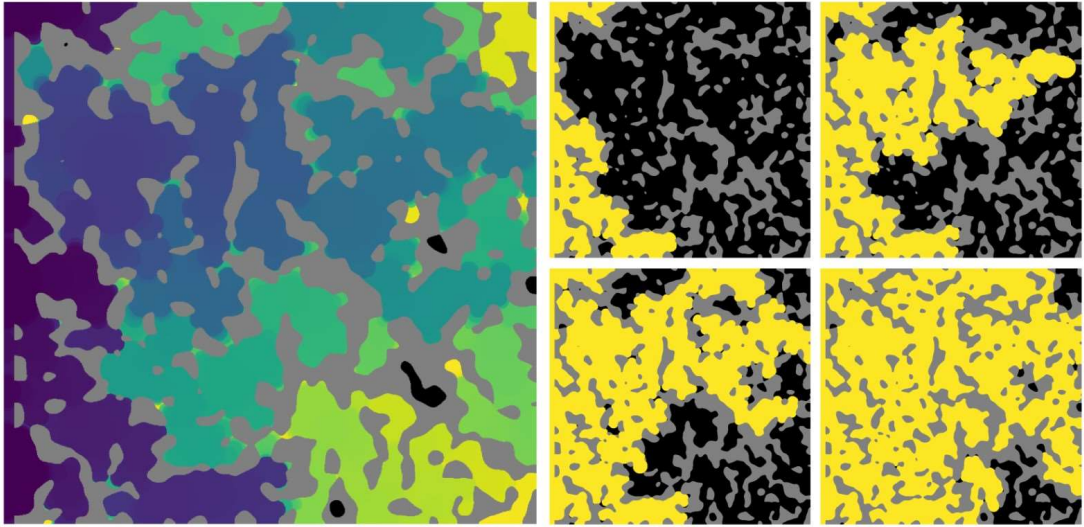

Figure 3: (left) Final result of invasion algorithm with each pixel/voxel labelled by the step at which it was first invaded. (right) Applying a threshold to the invasion image show at left produces a binary image of invaded (yellow) vs uninvaded (black) pore space. Invasion pattern at steps 500, 1500, 2000 and 3000 are shown. In all cases grey represents the solid phase.

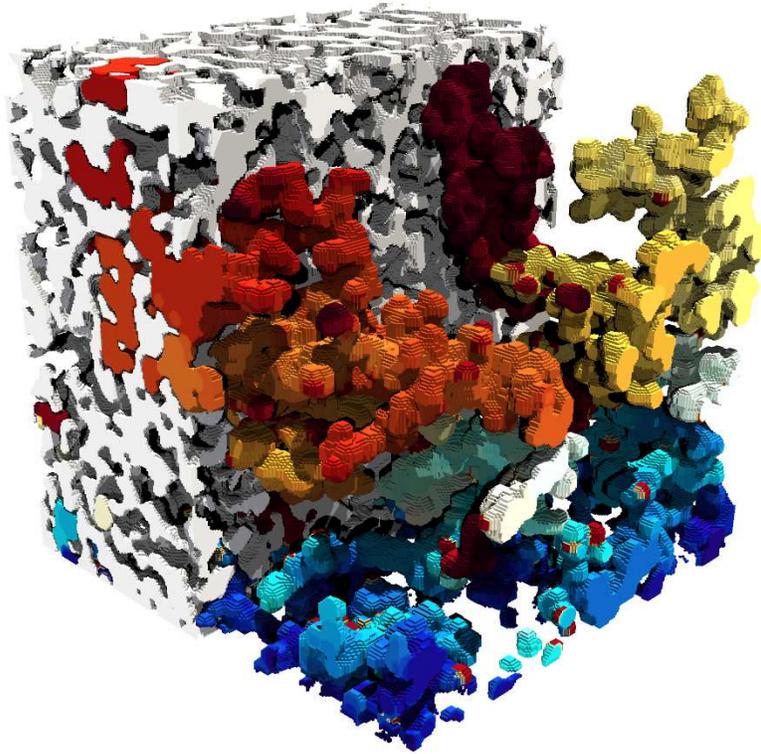

**Figure 4: Image of invasion configuration into a 3D domain of $250^3$ artificially generated porous media (blobs) with a porosity of 60%, with the solid phase indicated by white voxels. Invasion occurred from the bottom and the first 3,000 steps are shown. The color corresponds to the invasion step at which the voxel was invaded. This simulation required approximately 10 minutes on a normal laptop computer.**

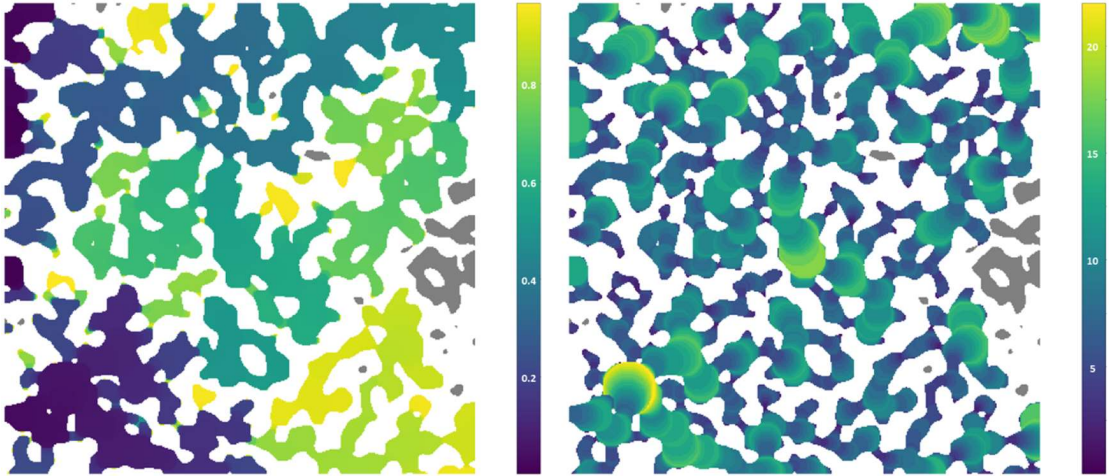

**Figure 5: (left)** The final result of the invasion algorithm with each voxel label according the image saturation at the point it was invaded. The color bar indicates the saturation. **(right)** During the invasion algorithm it is possible to insert sphere sizes instead of sequence number, so the color bar corresponds to the size of sphere that was inserted at each location.

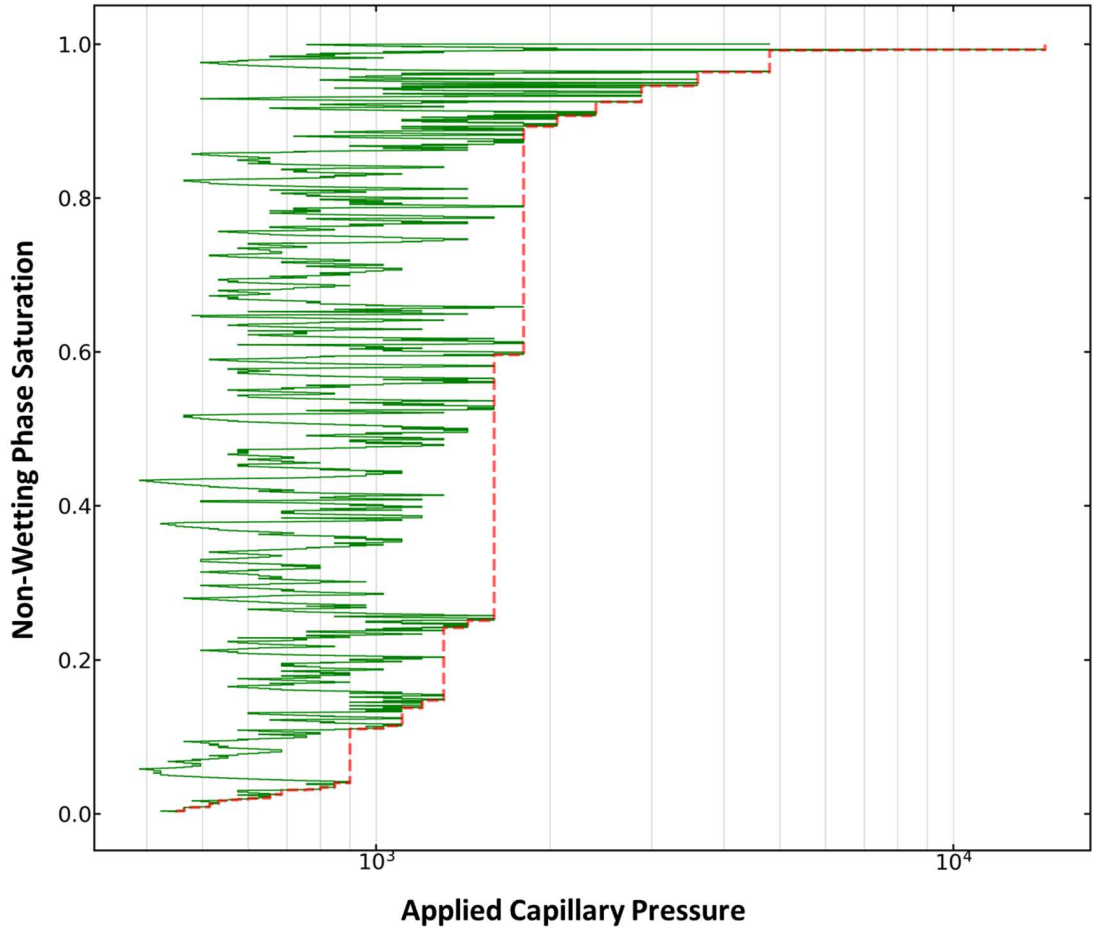

**Figure 6: Non-wetting phase saturation vs applied pressure graph. The red line was obtained by morphological image opening, and the green line was obtained the proposed algorithm.**

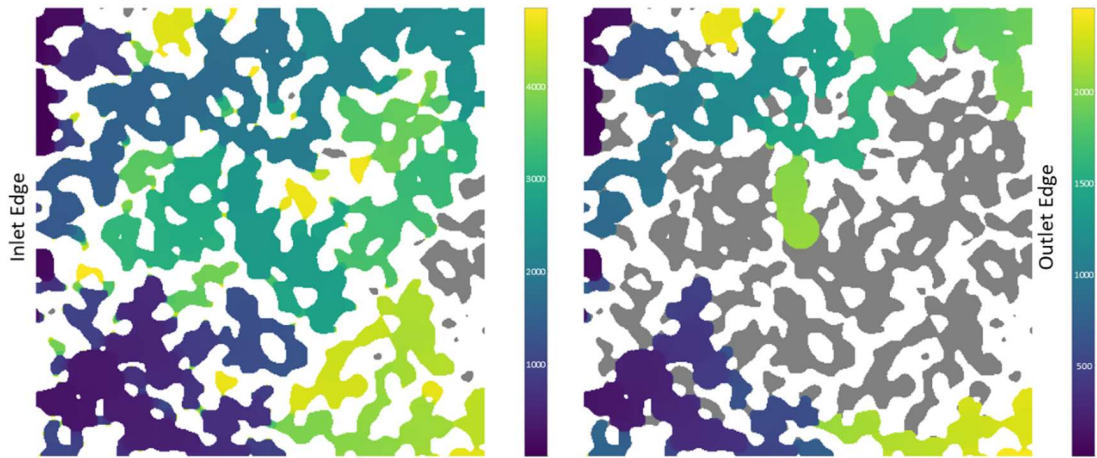

Figure 7: (left) Final result of invasion algorithm with each pixel/voxel labelled by the step at which it was first invaded. (right) Same simulation after applying a post-processing step to identify trapped voxels. Uninvaded voxels are indicated by grey. In the left image the only grey voxels are due inaccessible or blind pores. In the right image a significant portion of the pore space is grey due to trapping.

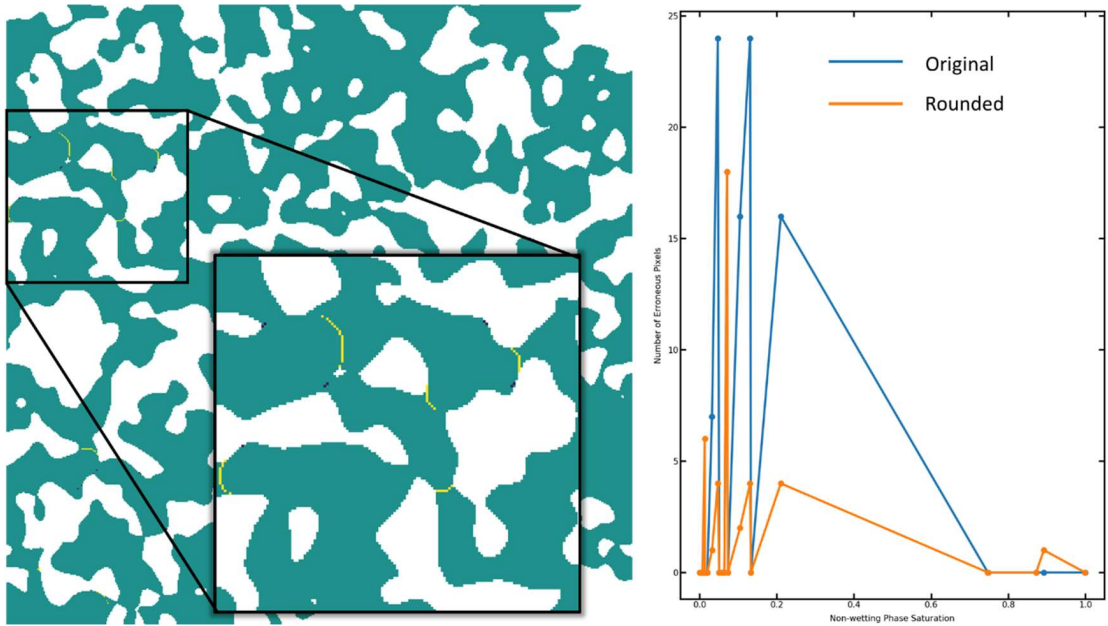

**Figure 8: Pixel-wise comparison of differences between MIO and ICE. Left shows the errors between the two fluid configurations at each saturation in the MIO image. Yellow indicates pixels invaded by MIO prior to ICE, purple indicates pixels invaded by IBIP prior to MIO. The inset shows a close-up of an area with several erroneous regions. Right shows the absolute error between the two methods in terms of number of pixels in disagreement at each saturation step. The maximum is 24 pixels in a $500^2$ image containing 250,000 pixels. The Rounded line is the disagreement if the saturation values are rounded to 4 decimal places, showing less mismatch.**

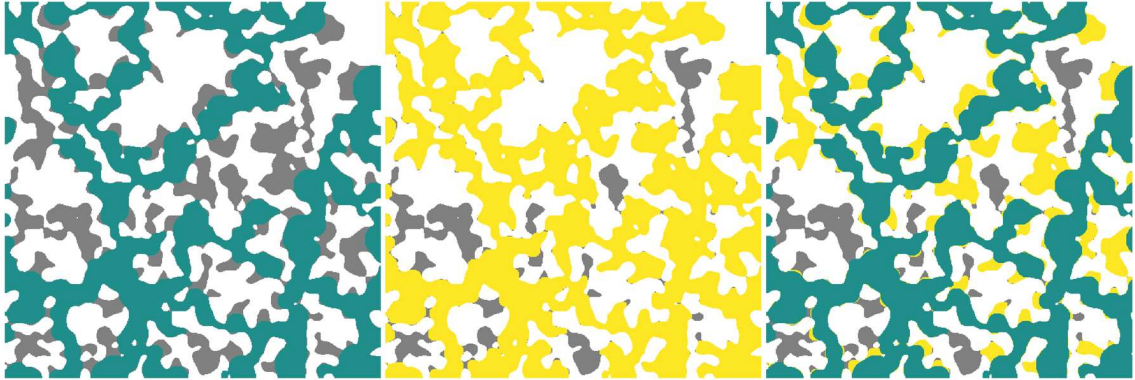

**Figure 9:** Comparison of trapping in IBIP (left) and MIO (center) with grey representing trapped defender and white representing solid. The image on the right shows the two results overlapped, so any visible yellow pixels are trapped phase incorrectly identified by the MIO method. Note that blind pores have been removed from the base image for clarity.

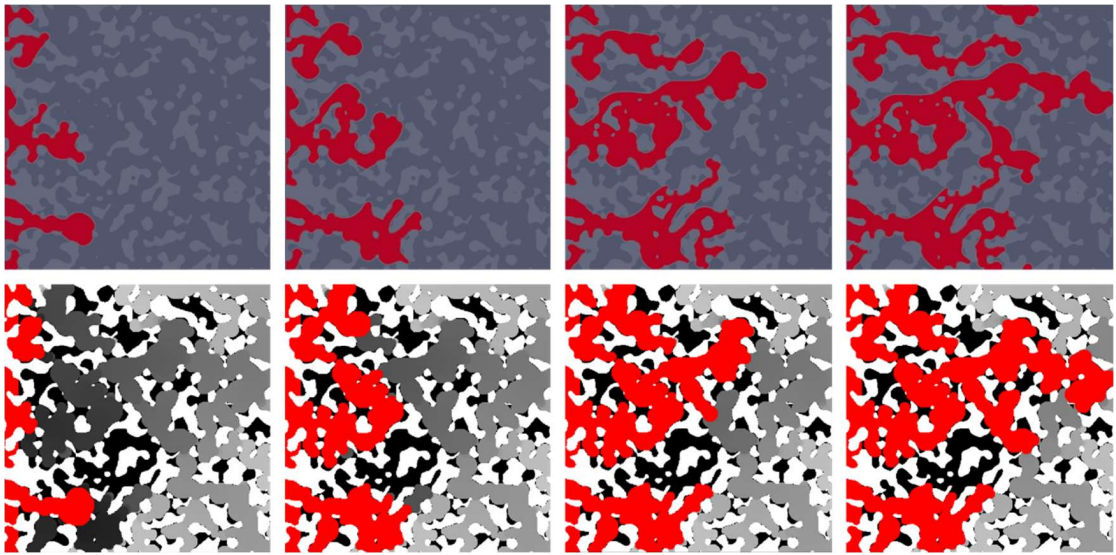

Figure 10: Comparison of 2-phase LBM results (top) with IBIP (bottom) on a 2D image of blobs at various levels of invasion up to the breakthrough point.